\documentclass[twocolumn, journal]{IEEEtran}
\IEEEoverridecommandlockouts

\usepackage{color}
\usepackage{bm}
\usepackage{graphicx}
\usepackage{amsmath}
\usepackage{amssymb}
\usepackage{algorithm}
\usepackage{algorithmic}
\usepackage{multirow}
\usepackage{booktabs}
\usepackage{array}
\usepackage{amsthm}
\usepackage{lipsum}
\usepackage{enumerate}
\usepackage{stfloats}
\usepackage{subfigure}
\usepackage{cases}
\usepackage{diagbox}
\usepackage{threeparttable}
\usepackage{cite}

\allowdisplaybreaks[4]

\title{\huge Beyond Diagonal Reconfigurable Intelligent Surfaces with Mutual Coupling: Modeling and Optimization}

\author{Hongyu Li,~\IEEEmembership{Student Member,~IEEE}, Shanpu Shen,~\IEEEmembership{Senior Member},
Matteo Nerini,~\IEEEmembership{Student Member,~IEEE},\\ Marco Di Renzo,~\IEEEmembership{Fellow,~IEEE}, and Bruno Clerckx,~\IEEEmembership{Fellow,~IEEE}
\thanks{Manuscript received 4 October 2023; revised 18 December 2023; accepted 28 January 2024. 
The work of M. Di Renzo was supported in part by the EC HE projects COVER-101086228, UNITE-101129618, INSTINCT-101139161, and the ANR projects PEPR-NF-YACARI 22-PEFT-0005, PASSIONATE ANR-23-CHR4-0003-01.
The associate editor coordinating the review of this letter and approving it for publication was José Cândido Silveira Santos Filho.}
\thanks{H. Li and M. Nerini are with the Department of Electrical and Electronic Engineering, Imperial College London, London SW7 2AZ, U.K. (e-mail:\{c.li21,~m.nerini20\}@imperial.ac.uk).}
\thanks{S. Shen is with the Department of Electrical Engineering and Electronics, University of Liverpool, Liverpool L69 3GJ, U.K. (email: Shanpu.Shen@liverpool.ac.uk).}
\thanks{M. Di Renzo is with Universit\'e Paris-Saclay, CNRS, CentraleSup\'elec, Laboratoire des Signaux et Syst\`emes, 3 Rue Joliot-Curie, 91192 Gif-sur-Yvette, France. (email: marco.di-renzo@universite-paris-saclay.fr)}
\thanks{B. Clerckx is with the Department of Electrical and Electronic Engineering, Imperial College London, London SW7 2AZ, U.K. and with Silicon Austria Labs (SAL), Graz A-8010, Austria (e-mail: b.clerckx@imperial.ac.uk; bruno.clerckx@silicon-austria.com).}}

\begin{document}

\maketitle
\thispagestyle{empty}
\begin{abstract}
This work studies the modeling and optimization of beyond diagonal reconfigurable intelligent surface (BD-RIS) aided wireless communication systems in the presence of mutual coupling among the RIS elements.
Specifically, we first derive the mutual coupling aware BD-RIS aided communication model using scattering and impedance parameters.
Based on the obtained communication model, we propose a general BD-RIS optimization algorithm applicable to different architectures of BD-RIS to maximize the channel gain. 
Numerical results validate the effectiveness of the proposed design and demonstrate that the larger the mutual coupling the larger the gain offered by BD-RIS over conventional diagonal RIS.
\end{abstract}


\vspace{-0.5 cm}

\section{Introduction}
\label{sec:intro}

\vspace{-0.1 cm}

Beyond diagonal reconfigurable intelligent surface (BD-RIS) has recently been introduced as a new advance in the context of RIS-aided communications \cite{wu2021intelligent}, which generalizes conventional RIS with diagonal impedance-controlled matrices and results in scattering matrices that are not diagonal. 
This is realized through reconfigurable inter-element connections among the RIS elements \cite{shen2021}, resulting in better controllability of the scattered waves to boost the rate and coverage \cite{li2023reconfigurable}.

Existing BD-RIS works mainly focus on modeling \cite{shen2021} and mode/architecture design \cite{zhang2022intelligent,li2023reconfigurable,li2022reconfigurable,nerini2023beyond}.
BD-RIS was first modeled in \cite{shen2021} using the scattering parameters, where single, group, and fully-connected architectures are introduced based on the circuit topology of the tunable RIS impedance network. 
Inspired by the intelligent omni-surface with enlarged coverage compared to conventional RIS \cite{zhang2022intelligent} and thanks to the structural flexibility introduced by the more general scattering matrices, BD-RIS with hybrid and multi-sector modes \cite{li2023reconfigurable} are proposed, which achieve full-space coverage with enhanced performance. 
To explore the best performance-complexity trade-off provided by BD-RIS, other special architectures, including BD-RIS with non-diagonal phase shift matrices \cite{li2022reconfigurable} and BD-RIS with tree- and forest-connected architectures \cite{nerini2023beyond} have been proposed.
However, all the existing BD-RIS works \cite{shen2021,li2023reconfigurable,zhang2022intelligent,li2022reconfigurable,nerini2023beyond} focus on idealized cases without considering the impact of mutual coupling among the RIS elements.

Mutual coupling is important and cannot be ignored in practice, given that BD-RIS architectures usually consist of numerous densely spaced elements within a limited aperture to increase the beamforming gain and to realize sophisticated wave transformations \cite{di2022communication}.
There are only limited works on conventional RIS analyzing the impact of mutual coupling \cite{di2022communication,gradoni2021end,qian2021mutual}. 
The general modeling of BD-RIS including the antenna mutual coupling and mismatching is introduced in \cite{shen2021}.
However, it is too complicated to explicitly understand the impact of mutual coupling. 
Furthermore, the optimization of BD-RIS aided wireless communication systems in the presence of mutual coupling has never been investigated.

Motivated by these considerations, in this work, we model, analyze, and optimize BD-RIS in the presence of mutual coupling. The contributions of this work are summarized as follows.
\textit{First}, we derive the BD-RIS aided wireless communication model, which captures the mutual coupling among the RIS elements. This is done by leveraging the equivalence between the scattering and impedance parameters. 
\textit{Second}, we propose a general and efficient optimization algorithm to maximize the channel gain for a BD-RIS aided single-input single-output (SISO) system, which can be applied to BD-RIS with single-, group-, and fully-connected architectures. 
\textit{Third}, we present simulation results to verify the effectiveness of the proposed algorithm and show the performance enhancement of BD-RIS with group/fully-connected architectures when taking into account the mutual coupling at the optimization stage.

\textit{Notations:}
$\Re\{\cdot\}$ and $\Im\{\cdot\}$ denote the real and imaginary parts of complex variables.
$\otimes$ denotes the Kronecker product.
$\mathsf{blkdiag}(\cdot)$ denotes a block-diagonal matrix.
$\mathsf{vec}(\cdot)$ denotes the vectorization.
$\overline{\mathsf{vec}}(\cdot)$ reshapes a vectorized matrix into the original matrix. 
$\angle(\cdot)$ returns the angle of complex variables.
$\mathbf{I}_M$ is an $M\times M$ identity matrix.
$[\mathbf{A}]_{i:i',j:j'}$ extracts a sub-matrix of $\mathbf{A}$ from $i$-th to $i'$-th rows and $j$-th to $j'$-th columns.

\vspace{-0.2 cm}

\section{RIS Aided Communication Model}
\label{sec:syst_mod}

In this section, we review the general RIS aided communication model derived in \cite{shen2021} and simplify it while still modeling the mutual coupling among the RIS elements. 

\vspace{-0.2 cm}

\subsection{General RIS Aided Communication Model}

We consider a BD-RIS aided multi-antenna system with an $N$-antenna transmitter, an $M$-element BD-RIS, and a $K$-antenna receiver.
The whole system is modeled as an $L$-port network with $L = N+M+K$, which can be characterized by a scattering matrix $\mathbf{S}\in\mathbb{C}^{L\times L}$ \cite{shen2021}.  
The matrix $\mathbf{S}$ can be formulated in terms of sub-matrices as $\mathbf{S} = [\mathbf{S}_{TT}, \mathbf{S}_{TI}, \mathbf{S}_{TR}; \mathbf{S}_{IT}, \mathbf{S}_{II}, \mathbf{S}_{IR}; \mathbf{S}_{RT}, \mathbf{S}_{RI}, \mathbf{S}_{RR}]$, where the diagonal sub-matrices refer to the scattering matrices that correspond to the transmitter, RIS, and receiver radiating elements. The off-diagonal sub-matrices refer to the transmission scattering matrices between the transmitter, RIS, and receiver.

\begin{figure}
    \centering
    \includegraphics[width = 0.48\textwidth]{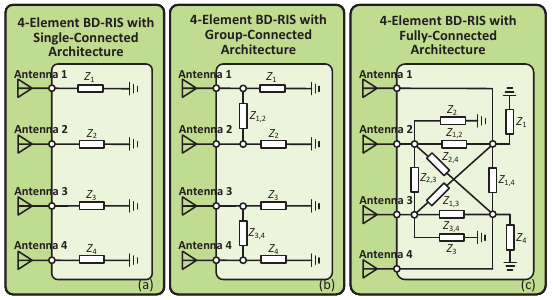}
    \caption{Schematic circuit topologies for a 4-element BD-RIS with (a) single-connected, (b) group-connected, and (c) fully-connected architectures.}
    \label{fig:architecture}\vspace{-0.5 cm}
\end{figure}

In addition, we assume that each transmit antenna is connected in series to a voltage source and a source impedance, yielding a source impedance matrix $\mathbf{Z}_T\in\mathbb{C}^{N\times N}$. Accordingly, the reflection coefficient matrix at the transmitter is $\mathbf{\Gamma}_T = (\mathbf{Z}_T+Z_0\mathbf{I}_N)^{-1}(\mathbf{Z}_T-Z_0\mathbf{I}_N)$, where $Z_0$ is the reference impedance used to calculate the scattering matrix.
Similarly, each receive antenna is connected to the ground through a load impedance, yielding a load impedance matrix $\mathbf{Z}_R\in\mathbb{C}^{K\times K}$ and a reflection coefficient matrix $\mathbf{\Gamma}_R = (\mathbf{Z}_R+Z_0\mathbf{I}_K)^{-1}(\mathbf{Z}_R-Z_0\mathbf{I}_K)$.
The $M$ scattering elements at the BD-RIS are connected to an $M$-port group-connected reconfigurable impedance network \cite{shen2021}, where the $M$ ports are uniformly divided into $G$ groups, each containing $\bar{M} = \frac{M}{G}$ ports connected to each other.  
Mathematically, BD-RIS with a group-connected architecture is characterized by a block-diagonal impedance matrix $\mathbf{Z}_I\in\mathbb{C}^{M\times M}$, i.e.,
\begin{equation}
    \label{eq:z_blkdiag}
    \mathbf{Z}_I = \mathsf{blkdiag}(\mathbf{Z}_{I,1},\ldots,\mathbf{Z}_{I,G}),
\end{equation}
where each block $\mathbf{Z}_{I,g}\in\mathbb{C}^{\bar{M}\times\bar{M}}$, $\forall g\in\mathcal{G}=\{1,\ldots,G\}$, is symmetric and purely imaginary for reciprocal and lossless reconfigurable impedance networks \cite{shen2021}, i.e.,
\begin{equation}
    \label{eq:z_cons}
    \mathbf{Z}_{I,g} = \mathbf{Z}_{I,g}^T, ~ \Re\{\mathbf{Z}_{I,g}\} = \mathbf{0}, \forall g\in\mathcal{G}.
\end{equation}
Specifically, the cases with $G = 1$ and $G=M$ refer to the fully- and single-connected architectures of BD-RIS \cite{shen2021}, where the corresponding impedance matrices become full matrices and diagonal matrices, respectively.
To facilitate understanding, we provide the schematic circuit topologies of BD-RIS with single-, group-, and fully-connected architectures in Fig. \ref{fig:architecture}.
The impedance matrix in (\ref{eq:z_cons}) results in a symmetric and unitary matrix $\mathbf{\Theta}\in\mathbb{C}^{M\times M}$ of reflection coefficients with 
\begin{equation}
    \label{eq:theta_z}
    \mathbf{\Theta} = (\mathbf{Z}_I+Z_0\mathbf{I}_M)^{-1}(\mathbf{Z}_I-Z_0\mathbf{I}_M).
\end{equation}

Using these definitions and multiport network theory based on the scattering parameters, the general RIS aided channel, $\mathbf{H}\in\mathbb{C}^{K\times N}$, which relates the voltages at the receiver ports with those at transmitter ports, is given by\cite{shen2021}
\begin{equation}
    \label{eq:general_channel}
    \mathbf{H} = (\mathbf{\Gamma}_R+\mathbf{I}_K)^{-1}\mathbf{T}_{RT}(\mathbf{I}_N+\mathbf{\Gamma}_T\mathbf{T}_{TT}+\mathbf{T}_{TT})^{-1},
\end{equation}
where $\mathbf{T}_{TT} \in\mathbb{C}^{N\times N}$ and $\mathbf{T}_{RT}\in\mathbb{C}^{K\times N}$ are sub-matrices of $\mathbf{T} = \mathbf{S}(\mathbf{I}_L-\mathbf{\Gamma}\mathbf{S})^{-1}\in\mathbb{C}^{L\times L}$ with $\mathbf{\Gamma} = \mathsf{blkdiag}(\mathbf{\Gamma}_T,\mathbf{\Theta},\mathbf{\Gamma}_R)$. 
Specifically, $\mathbf{T}_{TT} = [\mathbf{T}]_{1:N,1:N}$ and $\mathbf{T}_{RT} = [\mathbf{T}]_{N+M+1:L,1:N}$.
The channel model in (\ref{eq:general_channel}) is general enough to include the impact of antenna mismatching and mutual coupling at the transmitter, BD-RIS, and receiver. However, it is too complicated to get insights on the role of BD-RIS, which motivates us to apply further simplifications.

\vspace{-0.3 cm}

\subsection{Mutual Coupling Aware RIS Aided Communication Model}

To simplify the general communication model (\ref{eq:general_channel}), we introduce the impedance matrix $\mathbf{Z}\in\mathbb{C}^{L\times L}$ for the whole $L$-port network, which can be formulated in terms of sub-matrices, as $\mathbf{Z} = [\mathbf{Z}_{TT}, \mathbf{Z}_{TI}, \mathbf{Z}_{TR}; \mathbf{Z}_{IT}, \mathbf{Z}_{II}, \mathbf{Z}_{IR}; \mathbf{Z}_{RT}, \mathbf{Z}_{RI}, \mathbf{Z}_{RR}]$. 
Then, we consider the following assumptions.

\textit{Assumption 1:} The source impedances at the transmitter and the load impedances at the receiver are equal to the reference impedance $Z_0$, i.e., $\mathbf{Z}_T = Z_0\mathbf{I}_N$ and $\mathbf{Z}_R = Z_0\mathbf{I}_K$, such that $\mathbf{\Gamma}_T = \mathbf{0}$ and $\mathbf{\Gamma}_R = \mathbf{0}$.
This corresponds to the best power matching at the transmitter and receiver, respectively. 

\textit{Assumption 2 (Unilateral Approximation\footnote{The accuracy of the unilateral approximation is discussed in \cite{ivrlavc2010toward}, to which interested readers are referred for further information.
}):} The distances between the transmitter and receiver, transmitter and BD-RIS, and BD-RIS and receiver are sufficiently large such that the links from the receiving devices to the transmitting devices are negligible, i.e., $\mathbf{Z}_{TI} \approx \mathbf{0}$, $\mathbf{Z}_{TR} \approx \mathbf{0}$, $\mathbf{Z}_{IR} \approx \mathbf{0}$.

\textit{Assumption 3:} The antennas at the transmitter and receiver are perfectly matched with no mutual coupling, such that $\mathbf{Z}_{TT} = Z_0\mathbf{I}_N$ and $\mathbf{Z}_{RR} = Z_0\mathbf{I}_K$.

Based on Assumptions 2 and 3, and the relationship $\mathbf{S} = (\mathbf{Z} +Z_0\mathbf{I}_L)^{-1}(\mathbf{Z}-Z_0\mathbf{I}_L)$, we obtain the following result. 

\textit{Result 1:} The transmission scattering matrices from the receiver to the transmitter are almost zero, i.e., $\mathbf{S}_{TI} \approx \mathbf{0}$, $\mathbf{S}_{TR} \approx \mathbf{0}$, $\mathbf{S}_{IR} \approx \mathbf{0}$. The scattering matrices at the antenna arrays of the transmitter and receiver are almost zero, i.e., $\mathbf{S}_{TT} \approx \mathbf{0}$, $\mathbf{S}_{RR} \approx \mathbf{0}$.

Applying Assumption 1 and Result 1, the expression in (\ref{eq:general_channel}) can be simplified as follows: 
\begin{equation}
    \label{eq:channel}
        \mathbf{H} \approx \mathbf{S}_{RT} + \mathbf{S}_{RI}(\mathbf{I}_M - \mathbf{\Theta}\mathbf{S}_{II})^{-1}\mathbf{\Theta}\mathbf{S}_{IT}.
\end{equation}
However, (\ref{eq:channel}) is still too complex, since the matrix $\mathbf{\Theta}$ of reflection coefficients appears inside and outside the inverse. 
To further simplify the expression in (\ref{eq:channel}), so as to facilitate the optimization of BD-RIS in the presence of mutual coupling, we 
leverage Assumptions 1-3 and the relationship between $\mathbf{S}$ and $\mathbf{Z}$, and obtain the following result. 

\textit{Result 2:}
The nonzero sub-matrices of $\mathbf{S}$ and those of $\mathbf{Z}$ are related to one another as follows:
\begin{subequations}
    \label{eq:s2z}
    \begin{align}
        \mathbf{S}_{RI} & = \frac{\mathbf{Z}_{RI}}{2Z_0}(\mathbf{I}_M - (\mathbf{Z}_{II} +Z_0\mathbf{I}_M)^{-1}(\mathbf{Z}_{II} - Z_0\mathbf{I}_M)),\\
        \mathbf{S}_{II} &= (\mathbf{Z}_{II} +Z_0\mathbf{I}_M)^{-1}(\mathbf{Z}_{II} - Z_0\mathbf{I}_M),\\
        \mathbf{S}_{IT} &= (\mathbf{Z}_{II} +Z_0\mathbf{I}_M)^{-1}\mathbf{Z}_{IT},\\
        \mathbf{S}_{RT} &= \frac{\mathbf{Z}_{RT}}{2Z_0}-\frac{\mathbf{Z}_{RI}}{2Z_0}(\mathbf{Z}_{II} +Z_0\mathbf{I}_M)^{-1}\mathbf{Z}_{IT}.    
    \end{align}
\end{subequations}

Plugging (\ref{eq:s2z}) and (\ref{eq:theta_z}) into (\ref{eq:channel}), we obtain a more tractable and convenient expression equivalent to (\ref{eq:channel}) as
\begin{equation}
    \label{eq:channel1}
    \mathbf{H} \approx \frac{1}{2Z_0}(\mathbf{Z}_{RT} - \mathbf{Z}_{RI}(\mathbf{Z}_{II} + \mathbf{Z}_I)^{-1}\mathbf{Z}_{IT}).
\end{equation}
Equation (\ref{eq:channel1}) is in accordance with the model in \cite{gradoni2021end}, which was derived using the impedance parameter analysis\footnote{The end-to-end model derived in \cite{gradoni2021end} relates the voltage at the receiver ports and the source voltage, which can be easily transformed into (\ref{eq:channel1}) using the relationship between the source voltage and the voltage at the transmitter.}. 
In (\ref{eq:channel1}), the role of the beyond-diagonal network of tunable impedances in BD-RIS and the impact of mutual coupling and mismatching at the RIS elements are explicitly visible. 
The physical meaning of each term in (\ref{eq:channel1}) is explained as follows. 
\begin{enumerate}[1)]
    \item $\mathbf{Z}_{RT}$, $\mathbf{Z}_{RI}$, and $\mathbf{Z}_{IT}$ refer to the channels from the transmitter to the receiver, from the BD-RIS to the receiver, and from the transmitter to the BD-RIS, respectively.
    \item $\mathbf{Z}_{II}$ characterizes the mismatching and mutual coupling at the RIS elements. Specifically, the diagonal entries of $\mathbf{Z}_{II}$ refer to the self impedance; the off-diagonal entries of $\mathbf{Z}_{II}$ account for the mutual coupling, which depends on the inter-element distance. Generally, the larger the inter-element distance, the smaller the mutual coupling. 
\end{enumerate}
With the communication model in (\ref{eq:channel1}) at hand, which explicitly accounts for the mutual coupling at the BD-RIS, in the following section, we aim to design the beyond-diagonal matrix $\mathbf{Z}_I$ of tunable impedances to maximize the system performance. 

\textit{Remark 1:} We clarify that we start from the general channel model derived in \cite{shen2021}, but go beyond that by deriving the new expression in (\ref{eq:channel}), which captures explicitly the mutual coupling among the RIS elements. 
For the first time, more importantly, we show the equivalence between the scattering parameter based channel in (\ref{eq:channel}) and the impedance parameter based channel in (\ref{eq:channel1}) derived in \cite{gradoni2021end}.
In addition, only the conventional RIS with single-connected architecture is considered in \cite{gradoni2021end}, i.e., $\mathbf{Z}_I$ is diagonal, while in this work we consider BD-RIS with $\mathbf{Z}_I$ not limited to being diagonal. 

\vspace{-0.2 cm}

\section{Mutual Coupling Aware Optimization}
\label{sec:opt}
For simplicity, we consider a SISO system and focus on the optimization of $\mathbf{Z}_I$, which is not limited to being diagonal in BD-RIS. 
The analysis of the multiple-antenna case is postponed to a future work due to the complexity of considering a joint optimization including the precoder at the transmitter, the beyond-diagonal matrix $\mathbf{Z}_I$ at the BD-RIS, and the combiner at the receiver.
The corresponding optimization problem to maximize the channel gain can be formulated as\footnote{Here, we assume that channel state information (CSI) is perfectly known at the transmitter to better focus on and highlight the impact of mutual coupling.}
\begin{equation}
    \label{eq:p0}
    \begin{aligned}
        \max_{\mathbf{Z}_I} ~ & |z_{RT} - \mathbf{z}_{RI}(\mathbf{Z}_{II} + \mathbf{Z}_{I})^{-1}\mathbf{z}_{IT}|^2~~
        \mathrm{s.t.} ~\text{(\ref{eq:z_blkdiag}), (\ref{eq:z_cons})},
    \end{aligned}
\end{equation}
where $z_{RT}\in\mathbb{C}$, $\mathbf{z}_{RI}\in\mathbb{C}^{1\times M}$, and $\mathbf{z}_{IT}\in\mathbb{C}^{M\times 1}$.
The main difficulties when solving the problem in (\ref{eq:p0}) are the following:
\begin{enumerate}[1)]
    \item The matrix $\mathbf{Z}_{I}$ appears in the inversion $(\mathbf{Z}_{II} + \mathbf{Z}_{I})^{-1}$. 
    \item The matrix $\mathbf{Z}_{I}$ needs to fulfill the constraints (\ref{eq:z_blkdiag}) and (\ref{eq:z_cons}), instead of being diagonal as in conventional RISs. 
\end{enumerate}
The first difficulty has been tackled in \cite{qian2021mutual} for diagonal RISs.
However, the optimization with mutual coupling among the RIS elements and the unique constraints of BD-RIS have never been investigated. 
To effectively solve problem (\ref{eq:p0}), we apply the idea introduced in \cite{qian2021mutual} and propose to iteratively optimize $\mathbf{Z}_{I}$ subject to the constraints (\ref{eq:z_blkdiag}) and (\ref{eq:z_cons}) until convergence. This results in the following optimization framework. 

\vspace{-0.2 cm}

\subsection{Optimization Framework}
The main idea of the iterative design in \cite{qian2021mutual} is to slightly modify the value of $\mathbf{Z}_{I}$ in each iteration, in order to increase the channel gain. 
Inspired by this approach, we first introduce an auxiliary variable $\mathbf{\Omega}\in\mathbb{C}^{M\times M}$ as a small increment to $\mathbf{Z}_I$ in each iteration. 
To facilitate the iterative design, we construct $\mathbf{\Omega}$ based on the following two properties. 

\textit{Property 1: }
$\mathbf{\Omega}$ is chosen in compliance with the mathematical structure of the impedance matrix $\mathbf{Z}_I$, such that the updated impedance matrix always satisfies the optimization constraints in each iteration. Specifically, we have
\begin{equation}
    \label{eq:omega_blkdiag}
    \mathbf{\Omega} = \mathsf{blkdiag}(\mathbf{\Omega}_1,\ldots,\mathbf{\Omega}_G), ~\mathbf{\Omega}_g = \mathbf{\Omega}_g^T, \forall g.
\end{equation}

\textit{Property 2:}
Each nonzero entry of $\mathbf{\Omega}$ is chosen sufficiently small, such that the convergence of the optimization framework is guaranteed \cite{qian2021mutual}. 
For ease of optimization, we assume that each nonzero entry of $\mathbf{\Omega}$ has the same amplitude but a different phase, i.e.,
\begin{equation}
    \label{eq:omega_cons}
    |[\mathbf{\Omega}_g]_{m,n}| = \delta, \forall g, \forall m, n \in\bar{\mathcal{M}} = \{1,\ldots,\bar{M}\},
\end{equation}
where $\delta$ controls the increment of $\mathbf{Z}_I$ in each iteration, which is explained in details in the following subsection.

Based on these two properties for $\mathbf{\Omega}$, the proposed iterative design is summarized by the following steps. 

\textit{Step 1:} At the $l$-th iteration, we optimize $\mathbf{\Omega}^l$, while keeping $\mathbf{Z}_I^l$ fixed, by solving the following problem:
\begin{equation}
    \label{eq:p1}
    \begin{aligned}
        \max_{\mathbf{\Omega}} ~ & |z_{RT} - \mathbf{z}_{RI}(\mathbf{Z}_{II} + \mathbf{Z}_{I}^l + \mathbf{\Omega})^{-1}\mathbf{z}_{IT}|^2~~
        \mathrm{s.t.} ~ \text{(\ref{eq:omega_blkdiag}), (\ref{eq:omega_cons})}.
    \end{aligned}
\end{equation}

\textit{Step 2:} Given $\mathbf{\Omega}^l$ as the optimal solution to the problem in (\ref{eq:p1}), the impedance matrix at the $(l+1)$-th iteration, i.e., $\mathbf{Z}_I^{l+1}$, is updated using the following rule:
\begin{equation}
    \label{eq:update_Z}
    \mathbf{Z}_I^{l+1} = \mathbf{Z}_I^l + j\Im\{\mathbf{\Omega}^l\},
\end{equation}
where we only use the imaginary part of $\mathbf{\Omega}^l$ to guarantee that the impedance matrix is purely imaginary. 

Then, $\mathbf{Z}_I$ is optimized by iteratively solving the problem in (\ref{eq:p1}) and updating  (\ref{eq:update_Z}) until convergence. 
The solution of the problem in (\ref{eq:p1}) is detailed next. 

\vspace{-0.3 cm}

\subsection{Solution to Problem (\ref{eq:p1})}
In (\ref{eq:p1}), instead of optimizing $\mathbf{Z}_I$ as in (\ref{eq:p0}), we optimize the small increment $\mathbf{\Omega}$, such that $\mathbf{\Omega}$ can be removed from the matrix inversion in the objective function. 
Specifically, we apply the Neumann series approximation \cite{qian2021mutual} to the matrix inversion, i.e., 
$(\mathbf{Z}_{II} + \mathbf{Z}_I^l + \mathbf{\Omega})^{-1}
\approx (\mathbf{Z}_{II} + \mathbf{Z}_I^l)^{-1} - (\mathbf{Z}_{II} + \mathbf{Z}_I^l)^{-1}\mathbf{\Omega}(\mathbf{Z}_{II} + \mathbf{Z}_I^l)^{-1}$,
which achieves a tight approximation provided that the condition $\delta\ll\frac{1}{\bar{M}\|(\mathbf{Z}_{II} + \mathbf{Z}_I^l)^{-1}\|_\infty}$ is fulfilled in each iteration\footnote{For simplicity, in this work, we fix the value of $\delta$ as $\delta\ll 1$ to achieve a tight Neumann series approximation.}. 
Accordingly, we have $|z_{RT} - \mathbf{z}_{RI}(\mathbf{Z}_{II} + \mathbf{Z}_{I}^l + \mathbf{\Omega})^{-1}\mathbf{z}_{IT}|^2 \approx |a^l + \mathbf{b}^l\mathbf{\Omega}\mathbf{c}^l|^2 = |a^l + \sum_{g\in\mathcal{G}}\mathbf{e}_g^l\mathsf{vec}(\mathbf{\Omega}_g)|^2$, where $a^l = z_{RT} - \mathbf{z}_{RI}(\mathbf{Z}_{II} + \mathbf{Z}_{I}^l)^{-1}\mathbf{z}_{IT}$, $\mathbf{b}^l = \mathbf{z}_{RI}(\mathbf{Z}_{II} + \mathbf{Z}_{I}^l)^{-1}$, $\mathbf{c}^l = (\mathbf{Z}_{II} + \mathbf{Z}_{I}^l)^{-1}\mathbf{z}_{IT}$, $\mathbf{b}_g^l = [\mathbf{b}^l]_{(g-1)\bar{M}+1:g\bar{M}}$, $\mathbf{c}_g^l = [\mathbf{c}^l]_{(g-1)\bar{M}+1:g\bar{M}}$, and $\mathbf{e}_g^l = (\mathbf{c}_g^l)^T\otimes\mathbf{b}_g^l$, $\forall g\in\mathcal{G}$.
As a result, the problem in (\ref{eq:p1}) becomes 
\begin{equation}
    \label{eq:p2}
    \begin{aligned}
        \max_{\mathbf{\Omega}_g,\forall g} ~ & \Big|a^l + \sum_{g\in\mathcal{G}}\mathbf{e}_g^l\mathsf{vec}(\mathbf{\Omega}_g)\Big|^2~~
        \mathrm{s.t.} ~\mathbf{\Omega}_g = \mathbf{\Omega}_g^T, \text{(\ref{eq:omega_cons})}.
    \end{aligned}
\end{equation}

The solution to the problem in (\ref{eq:p2}) is not straightforward due to the symmetric and constant modulus constraints of $\mathbf{\Omega}_g, \forall g$. 
Fortunately, the symmetric constraint of each $\mathbf{\Omega}_g$ implies that we do not need to optimize the whole matrices $\mathbf{\Omega}_g, \forall g$.
In other words, instead of optimizing the symmetric matrices $\mathbf{\Omega}_g, \forall g$ with $\bar{M}^2$ variables, we can optimize $\bar{M}$ diagonal variables and $\frac{\bar{M}(\bar{M}-1)}{2}$ lower-triangular (or upper-triangular) variables of $\mathbf{\Omega}_g$, and reconstruct them to obtain $\mathbf{\Omega}_g, \forall g$. 
Mathematically, this transformation can be done by introducing the following notations:
\begin{enumerate}[1)]
    \item A column vector $\bm{\omega}_g\in\mathbb{C}^{\frac{\bar{M}(\bar{M}+1)}{2}\times 1}$, which contains the diagonal and lower-triangular entries of $\mathbf{\Omega}_g$.
    \item A binary matrix $\mathbf{P}\in\{0,1\}^{\bar{M}^2\times\frac{\bar{M}(\bar{M}+1)}{2}}$, which maps $\bm{\omega}_g$ into $\mathsf{vec}(\mathbf{\Omega}_g), \forall g$.
\end{enumerate}
Specifically, there is only one nonzero entry in each row of the binary matrix $\mathbf{P}$, which is defined as 
\begin{equation}
    \label{eq:p}
    [\mathbf{P}]_{\bar{M}(m-1)+n,k} = \begin{cases}
        1, & k = \frac{m(m-1)}{2}+n ~\text{and}~ 1 \le n \le m,\\
        1, & k = \frac{n(n-1)}{2}+m ~\text{and}~ m<n\le \bar{M},\\
        0, & \text{otherwise},
    \end{cases}
\end{equation}
where $\forall m, n \in\bar{\mathcal{M}}$. 
Accordingly, we obtain the relationship 
$\mathsf{vec}(\mathbf{\Omega}_g) = \mathbf{P}\bm{\omega}_g, \forall g$,
and the problem in (\ref{eq:p2}) becomes
\begin{equation}
    \label{eq:p3}
    \begin{aligned}
        \max_{\bm{\omega}_g,\forall g} ~ & \Big|a^l + \sum_{g\in\mathcal{G}} \mathbf{e}_g^l\mathbf{P}\bm{\omega}_g\Big|^2~~
        \mathrm{s.t.} ~|[\bm{\omega}_g]_i| = \delta, \forall g, \forall i\in\ddot{\mathcal{M}}, 
    \end{aligned}
\end{equation}
where $\ddot{\mathcal{M}}\in\{1,\ldots,\frac{\bar{M}(\bar{M}+1)}{2}\}$.
Applying the triangle inequality to the objective function in (\ref{eq:p3}), we have
$|a^l + \sum_{g\in\mathcal{G}} \mathbf{e}_g^l\mathbf{P}\bm{\omega}_g| \le |a^l| + \delta\sum_{g\in\mathcal{G}}\sum_{i\in\ddot{\mathcal{M}}} |[\mathbf{e}_g^l\mathbf{P}]_i|$,
where the equality is achieved by rotating each $[\mathbf{e}_g^l\mathbf{P}]_i$ such that the resulting $[\mathbf{e}_g^l\mathbf{P}]_i[\bm{\omega}_g]_i$, $\forall i$, are collinear to $a^l$ on the complex plane. 
Therefore, the optimal solution of $\bm{\omega}_g$, $\forall g$ in the $l$-th iteration can be determined element-by-element as 
\begin{equation}
    \label{eq:omega}
    [\bm{\omega}_g^l]_i = \delta\exp(j(\angle a^l - \angle [\mathbf{e}_g^l\mathbf{P}]_i)), \forall i.
\end{equation}
With the solution in (\ref{eq:omega}), we can reconstruct $\mathbf{\Omega}_g$ in the $l$-th iteration by $\mathbf{\Omega}_g^l = \overline{\mathsf{vec}}(\mathbf{P}\bm{\omega}_g^l)$, $\forall g$.

\textit{Remark 2:} The approach proposed in this letter is different from that in \cite{qian2021mutual} from two perspectives. \textit{First}, we introduce the auxiliary variable $\mathbf{\Omega}$ with the unique constraint in (\ref{eq:omega_blkdiag}) based on BD-RIS architectures, instead of being diagonal as in conventional RISs. \textit{Second}, we introduce the binary matrix $\mathbf{P}$ to deal with the symmetric constraint of $\mathbf{\Omega}$ and to further facilitate the optimization.

\vspace{-0.2 cm}

\subsection{Summary and Analysis}
\subsubsection{Algorithm} 
The complete algorithm for solving the problem in (\ref{eq:p0}) corresponds to the following steps:
\begin{enumerate}[S1:]
    \item Initialize $\mathbf{Z}_I^0$ by solving the problem in (\ref{eq:p0}) considering the conventional diagonal RIS with no mutual coupling, $\mathbf{\Omega}^0 = \mathbf{0}$, and $l=0$. Calculate $a^0$, $\mathbf{b}^0$, and $\mathbf{c}^0$. 
    \item Update $l = l +1$.
    \item Update $\mathbf{Z}_I^l$ using (\ref{eq:update_Z}), $a^l$, $\mathbf{b}^l$, and $\mathbf{c}^l$. 
    \item Update $\bm{\omega}_g^l$, $\forall g$ using (\ref{eq:omega}), and $\mathbf{\Omega}_g^l = \overline{\mathsf{vec}}(\mathbf{P}\bm{\omega}_g^l)$, $\forall g$.
    \item Calculate $C^l = |a^{l} + \mathbf{b}^{l}j\Im\{\mathbf{\Omega}^{l}\}\mathbf{c}^{l}|^2$.
    \item Repeat S2-S5 until the value of $C^l$ converges. 
\end{enumerate}

\subsubsection{Complexity} The complexity of the proposed algorithm mainly comes from the matrix inversion operation for updating $a^l$, $\mathbf{b}^l$, and $\mathbf{c}^l$, which requires $\mathcal{O}(M^3)$ complex multiplications. Therefore, the total complexity is $\mathcal{O}(IM^3)$, where $I$ denotes the number of iterations to ensure convergence. 

\subsubsection{Convergence} The convergence of the proposed algorithm is theoretically guaranteed by appropriately setting the value of $\delta$. 
Specifically, we have the relationship
$C^{l} = |a^{l} + \mathbf{b}^{l}j\Im\{\mathbf{\Omega}^{l}\}\mathbf{c}^{l}|^2 = |a^l + j\sum_{g\in\mathcal{G}}\mathbf{e}_g^l\mathbf{P}\Im\{\bm{\omega}_g\}|^2
\overset{\text{(a)}}{=} |j\delta\sum_{g}\sum_{i}|[\mathbf{e}_g^l\mathbf{P}]_i|\sin(\angle a^l - \angle[\mathbf{e}_g^l\mathbf{P}]_i)\cos(\angle a^l - \angle[\mathbf{e}_g^l\mathbf{P}]_i) + |a^l| + \delta\sum_{g}\sum_{i}|[\mathbf{e}_g^l\mathbf{P}]_i|\sin^2(\angle a^l - \angle[\mathbf{e}_g^l\mathbf{P}]_i)|^2
\ge |a^l|^2 = |z_{RT} - \mathbf{z}_{RI}(\mathbf{Z}_{II}+\mathbf{Z}_I^{l})^{-1}\mathbf{z}_{IT}|^2
= |z_{RT} - \mathbf{z}_{RI}(\mathbf{Z}_{II}+\mathbf{Z}_I^{l-1}+j\Im\{\mathbf{\Omega}^{l-1}\})^{-1}\mathbf{z}_{IT}|^2
\overset{\text{(b)}}{\approx} |a^{l-1} + \mathbf{b}^{l-1}j\Im\{\mathbf{\Omega}^{l-1}\}\mathbf{c}^{l-1}|^2 = C^{l-1}$,
where (a) follows by plugging (\ref{eq:omega}) into $C^l$; (b) follows from the Neumann series approximation.
Therefore, we have $C^{l}\ge|a^l|^2\approx C^{l-1}$, which proves that the objective function $C^l$ is monotonically non-decreasing after each iteration.
Therefore, we conclude that the proposed algorithm converges since the value of the objective function $C^l$ is bounded from above. 

\vspace{-0.4 cm}

\section{Performance Evaluation}
\label{sec:simulation}

In this section, we perform simulation results to evaluate the performance of the proposed algorithm.  
The simulation parameters are the same as in \cite{qian2021mutual} and are summarized as follows. 
Free-space propagation is assumed.
The transmitter and receiver are located at (5,-5,3) and (5,5,1), respectively.
BD-RIS is located on the $y$-$z$ plane and is centered at (0,0,0). 
All the radiating elements are thin wire dipoles that are parallel to the $z$-axis and have radius $r = \frac{\lambda}{500}$ and length $\iota = \frac{\lambda}{32}$, where $\lambda = \frac{c}{f}$ denotes the wavelength with frequency $f = 28$ GHz. 
Thus, the mutual impedance between any two dipoles $a$ and $b$ with center-point coordinates $(a_x,a_y,a_z)$ and $(b_x,b_y,b_z)$ (and the self impedance at any dipole if $a=b$) is computed as \cite{gradoni2021end}
\begin{equation}
    \label{eq:z_model}
    \begin{aligned}
    z_{a,b} &= \int_{a_z-\frac{\iota}{2}}^{a_z+\frac{\iota}{2}}\int_{b_z-\frac{\iota}{2}}^{b_z+\frac{\iota}{2}}  \frac{j\eta_0}{4\pi\kappa_0}\bigg(\frac{(a_z'-b_z')^2}{d_{a,b}^2}\big(\frac{3}{d_{a,b}^2}+\frac{j 3\kappa_0}{d_{a,b}} \\
    &~ -\kappa_0^2\big)- \frac{j\kappa_0+d_{a,b}^{-1}}{d_{a,b}}+\kappa_0^2\bigg)\frac{\exp(-j\kappa_0d_{a,b})}{d_{a,b}}\\
    &~\times\frac{\sin(\kappa_0(\frac{\iota}{2}-|b_z'-b_z|))\sin(\kappa_0(\frac{\iota}{2}-|a_z'-a_z|))}{\sin^2(\kappa_0\frac{\iota}{2})}db_z'da_z', 
    \end{aligned}
\end{equation}
where $\eta_0 = 377$ $\Omega$ denotes the free-space impedance, $\kappa_0 = \frac{2\pi}{\lambda}$ is the wavenumber, and $d_{a,b}$ denotes the distance between the dipoles $a$ and $b$, i.e., 
$d_{a,b} = \sqrt{(d_{a,b}^{x,y})^2 + (a_z'-b_z')^2}$, $d_{a,b}^{x,y} = \sqrt{(a_x-b_x)^2+(a_y-b_y)^2}$ for $a\ne b$ and $d_{a,b}^{x,y} = r$ for $a=b$.
The elements of $\mathbf{z}_{RI}$, $\mathbf{Z}_{II}$, and $\mathbf{z}_{IT}$ are computed using (\ref{eq:z_model}), and we set $z_{RT} = 0$ to focus on the role of BD-RIS.

\begin{figure}
    \centering
    \includegraphics[width = 0.48\textwidth]{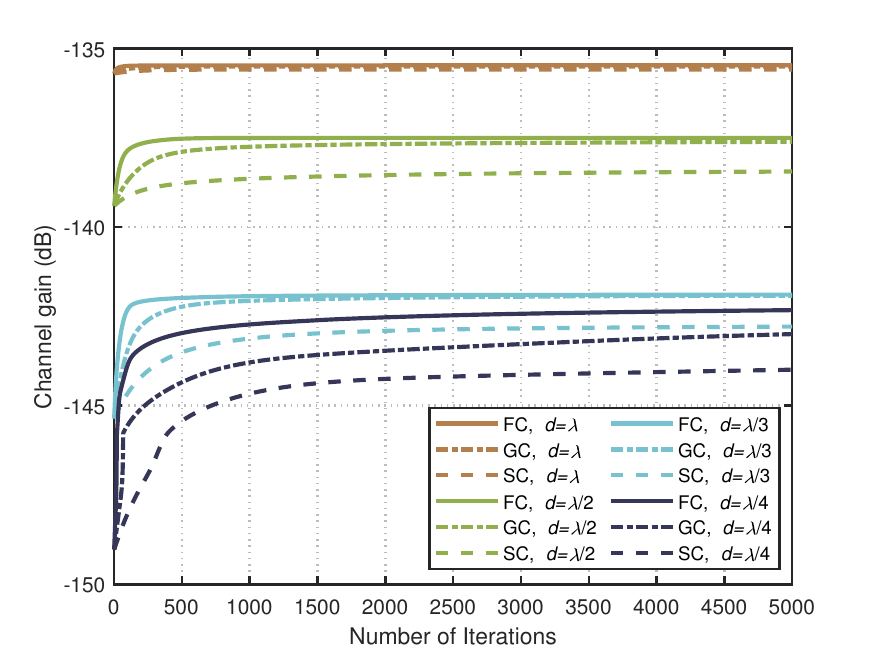}
    \vspace{-0.4 cm}
    \caption{Channel gain versus the number of iterations ($M = 16$).}
    \label{fig:channel_gain_Iter}\vspace{-0.48 cm}
\end{figure}

\begin{figure}
    \centering
    \includegraphics[width = 0.48\textwidth]{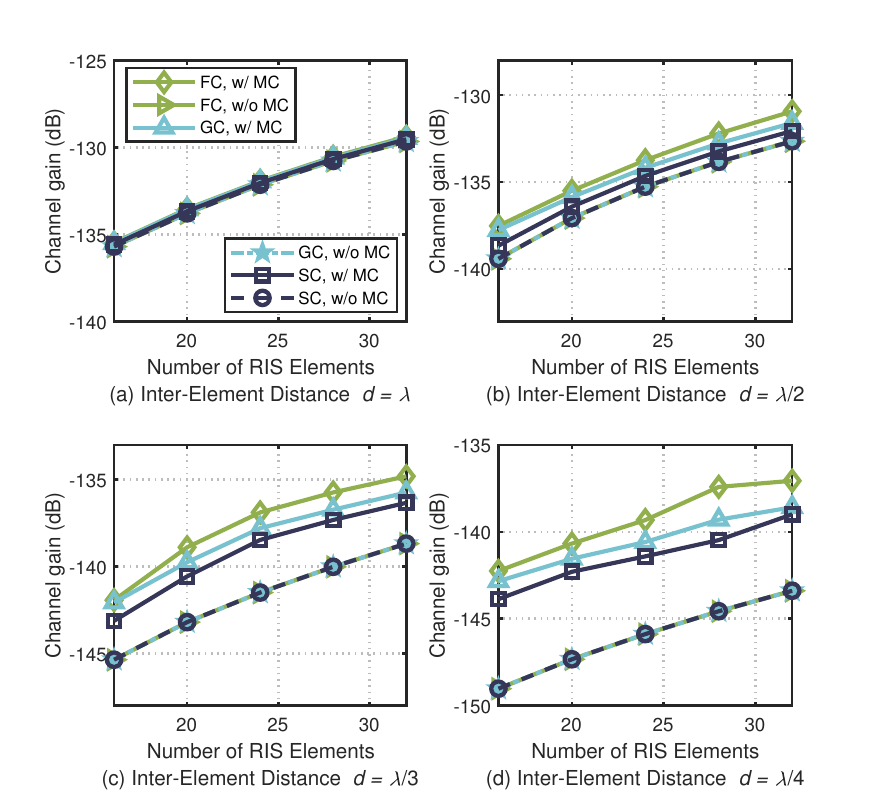}
    \vspace{-0.4 cm}
    \caption{Channel gain versus the number of BD-RIS elements.}
    \label{fig:channel_gain_M}\vspace{-0.6 cm}
\end{figure}

Fig. \ref{fig:channel_gain_Iter} evaluates the convergence of the proposed algorithm for different values of the inter-element distance $d$ and for different BD-RIS architectures. The schemes marked as ``FC/GC/SC'' refer to the performance achieved by BD-RIS with fully/group/single-connected architectures. 
We set $\bar{M} = 4$ for the group-connected case.
The value of $\delta$ is set to $6\times 10^{-4}$ to ensure convergence. 
We observe that the proposed algorithm converges, which verifies our theoretical derivation. 
In addition, we observe that, when the number of RIS elements is fixed, the end-to-end channel gain decreases when decreasing the inter-element distance due to the increased mutual coupling. When the aperture (size) of the RIS is kept fixed, on the other hand, a smaller inter-element distance enables the deployment of more RIS elements within the surface, which results in better beamforming gains provided that the mutual coupling is considered in the RIS design \cite{qian2021mutual}.

Fig. \ref{fig:channel_gain_M} illustrates the channel gain of BD-RIS with different architectures and different inter-element distances. 
In Fig. \ref{fig:channel_gain_M}, the schemes marked as ``w/ MC'' are obtained by computing $\mathbf{Z}_I$ with $\mathbf{Z}_{II}$ using (\ref{eq:z_model}); the schemes marked as ``w/o MC'' are obtained by setting the off-diagonal entries of $\mathbf{Z}_{II}$ to zero. In both cases, the channel gain is $|\mathbf{z}_{RI}(\mathbf{Z}_{II} + \mathbf{Z}_I)^{-1}\mathbf{z}_{IT}|^2$.
We have the following observations. 
\textit{First}, ignoring the mutual coupling when designing $\mathbf{Z}_I$, BD-RIS with single/group/fully-connected architectures achieves exactly the same performance, which is consistent with the conclusion in \cite{shen2021}. 
\textit{Second}, by taking into account the mutual coupling when designing $\mathbf{Z}_I$, BD-RIS with different architectures achieves better performance than the ``w/o MC'' schemes. 
\textit{Third}, the performance gap between fully/group-connected BD-RIS and conventional RIS increases when decreasing the inter-element distance. 
This can be attributed to the impact of mutual coupling, which becomes more prominent when decreasing the inter-element distance. This, in fact, results in larger values for the off-diagonal entries of $\mathbf{Z}_{II}$, which is better exploited by BD-RIS architectures for performance improvement. 

\vspace{-0.4 cm}

\section{Conclusion}
\label{sec:conclusion}

\vspace{-0.1 cm}

In this work, we studied the modeling and optimization of BD-RIS aided wireless communication systems in the presence of mutual coupling among the RIS elements. 
Specifically, we first derived a mutual coupling aware BD-RIS aided wireless communication model based on the scattering and impedance parameters, proving their equivalence.
We then proposed a general algorithm to maximize the channel gain of BD-RIS for SISO systems. 
We finally illustrated simulation results to analyze the effectiveness of the proposed design and the impact of mutual coupling on BD-RIS architectures. The numerical results showed that the larger the mutual coupling the larger the gain offered by BD-RIS.  

\vspace{-0.4 cm}

\bibliographystyle{IEEEtran}
\bibliography{refs}

\end{document}